\documentclass[twocolumn]{aastex701}
\usepackage{natbib}
\usepackage{amsmath}
\usepackage{graphicx}
\usepackage{subfigure}
\usepackage{float}
\usepackage{epsfig}
\usepackage{booktabs}
\usepackage{CJK}
\usepackage{verbatim}
\usepackage[normalem]{ulem}
\usepackage{xcolor}

\renewcommand{\d}[1]{\ensuremath{\operatorname{d}\!{#1}}}
\newcommand{\cntextsc}[1]
{\begin{CJK*}{UTF8}{gbsn}#1\end{CJK*}}

\shorttitle{Onset of Inside-Out Planet Formation}
\begin{document}
\begin{CJK*}{UTF8}{gkai}

\title{Inside-Out Planet Formation. VIII.\\Onset of Planet Formation and the Transition Disk Phase}

\author[0000-0003-3201-4549]{Xiao Hu (\cntextsc{胡晓})}
\affiliation{Department of Astronomy, University of Florida, Gainesville, FL 32608, USA}
\affiliation{Dept. of Astronomy, University of Virginia, Charlottesville, Virginia 22904, USA}
\email[show]{xiao.hu.astro@gmail.com}  

\author[0000-0002-3389-9142]{Jonathan C. Tan}
\affiliation{Dept. of Astronomy, University of Virginia, Charlottesville, Virginia 22904, USA}
\affiliation{Dept. of Space, Earth \& Environment, Chalmers University of Technology, Gothenburg, Sweden}
\email[show]{jctan.astro@gmail.com}

\begin{abstract}
Inside-Out Planet Formation (IOPF) is a theory of {\it in situ}
formation via pebble accretion of close-in Earth to Super-Earth mass
planets at the pressure maximum associated with the dead zone inner
boundary (DZIB), whose location is set initially by thermal ionization of alkali
metals at $\sim1,200\:$K. With midplane disk temperatures determined
by viscous accretional heating, the radial location of the DZIB depends on the accretion rate of the disk. 
Here, we investigate the
ability of pebbles to be trapped at the DZIB as a function of the
accretion rate and pebble size. We discuss the conditions that are needed for pebble trapping to become efficient when the accretion rate drops to
$\sim10^{-9}\:M_\odot\:{\rm yr}^{-1}$ and the resulting DZIB is at $\sim 0.1\:$au, which is the expected evolutionary phase of the disk at the onset of IOPF.
This provides an important
boundary condition for IOPF theory, i.e., the properties of pebbles 
when planet formation begins. We find for our fiducial model that typical pebble sizes of $\sim0.5\:$mm are needed for pebble trapping to first become efficient at DZIBs near 0.1~au.
This model may also provide
an explanation for the first emergence of the transition disk phase in
protoplanetary disks with accretion rates of
$\sim10^{-9}\:M_\odot\:{\rm yr}^{-1}$.
\end{abstract}

\section{Introduction}

Among all the exoplanets discovered by {\it Kepler}, a significant
portion of them are Earth to Super-Earth sized in multi-planet
systems on aligned orbits that are smaller than that of Mercury. These ``systems with tightly-packed inner planets'' (STIPs) may reveal a dominant mode of planet formation in our Galaxy. There are currently two main
scenarios to explain STIP formation: (1) migration of planets
formed in the outer disk \citep[e.g.,][]{2012ARA&A..50..211K,
  2013A&A...553L...2C,2014IAUS..299..360C,Bitsch2015,Emsenhuber2021}; and (2) formation {\it in
  situ} \citep[e.g.,][]{2012ApJ.751.158,2013ApJ.775.53,2013MNRAS.431.3444C,CT14}.

It has been noted that the migration scenario can have difficulties to concentrate
planets close enough to match observed STIPs
\citep{2010MNRAS.401.1691M}. Also, migrating planets tend to become
trapped in orbits of low-order mean motion resonances, which is not a
particular feature of STIPs
\citep{2014prpl.conf..667B,2014ApJ...790..146F}. The discovery that
STIPs planets have masses and sizes consistent with rocky cores (when
revealed after evaporation of their H/He atmospheres) \citep{Rogers2021}, also suggests a formation location inside the water ice line and thus relatively close to the host star.

The first proposed {\it in situ} formation models started with high
concentrations of protoplanets inside 1~au, distributed with a power
law radial profile, e.g., scaled from the minimum mass solar
nebula. These protoplanets then go through dynamical collisional
growth \citep{2012ApJ.751.158,2013ApJ.775.53}. However, the initial
condition of such models may be subject to some difficulties, such as
early triggering of gravitational instability
\citep{2014MNRAS.440L..11R,2014ApJ...795L..15S} \citep[see
  also,][]{2013MNRAS.431.3444C}.  In addition, the dynamic effects
of gas on protoplanets during their growth tends to produce planets
with masses decreasing steeply with orbital radius, which is different
from the relatively flat scaling seen in STIPs
\citep{2015A&A...578A..36O}.

Inside-Out Planet Formation (IOPF) \citep[][hereafter Paper I]{CT14} is a different type of {\it in situ} formation
model, involving the sequential formation of planets fed by pebble
accretion with pebbles supplied from the outer disk and assumed to be
trapped at the pressure maximum associated with the dead zone inner
boundary (DZIB) with an inner magneto-rotational instability (MRI)-active region \citep{balbus91}. In the IOPF model, the DZIB is set initially by thermal ionization of alkali metals and/or thermionic emission from dust grains at temperatures of about 1,200~K that provide sufficient ionization to activate the MRI \citep[][hereafter IOPF Paper V]{Mohanty2018} \citep[see also][]{Jankovic2021,Jankovic2022}.
Once the first planet grows to be massive enough to open a shallow gap in the disk, then the local gas density decreases, the ionization fraction increases, and the MRI spreads outwards, causing the DZIB to retreat. However, the pressure maximum still exists, likely just a few tens of Hill radii away from the first planet \citep[][hereafter IOPF Paper III]{Hu2016}. The process of pebble trapping at the pressure maximum then repeats, and the above planet formation process repeats, thus building up a system of close-in planets sequentially from the inside-out. 

A key feature of the IOPF model is that there is a characteristic planetary core mass of a few Earth masses, which is set by the condition needed for gap opening in the disk, and that these cores are built of volatile poor material. Both of these predictions were later confirmed by the analysis of \citep{Rogers2021}. Note, also that these planetary cores are expected to be able to accrete a primordial H/He atmosphere with a few percent of the total planetary mass, which causes them to appear as Sub-Neptunes. Only the planets that later lose this atmosphere, expected to be via photoevaporation, then appear as Super-Earths.

Another feature of the IOPF model is a high flux of pebbles delivered to the inner disk, which, via astrochemical modeling, are predicted to deliver large amounts of water ice that becomes gas phase water where disk temperatures reach $\gtrsim 170\:$K, with implications for the C/O ratio of primordial atmospheres that form via gas accretion \citep[][hereafter IOPF Paper VII]{CevallosSoto22}. This prediction has been confirmed by recent {\it James Webb Space Telescope (JWST)} observations of protoplanetary disks \citep{Banzatti22}.

Concentrating pebbles at the DZIB is the first step of IOPF and the
status of the disk at this stage is crucial for determining the locations
and masses of the first and subsequent planets. \citet{Hu2018}
(hereafter IOPF Paper IV) studied pebble evolution in disks extending out to 30~au
that feed inner disk regions and explored disk parameters such as
accretion rate, $\dot{m}$, and viscosity in the DZIB region, $\alpha$,
finding that to match the locations and masses of observed planets
requires accretion rates $\dot{m} \sim 10^{-9}\:M_\odot\:{\rm
  yr}^{-1}$ and DZIB viscosities $\alpha\sim10^{-4}$.

The location of dust and pebbles in protoplanetary disks under the
influence of already formed single or multiple planets has been
studied extensively
\citep[e.g.,][]{zhu2012,Pinilla2016,BZ2016,Dong2017}.  The disk
profile under perturbation in these disks is well-defined by
planet-disk dynamics. This is not the case for the initial DZIB in the
IOPF model that is yet to form a planet. Observationally, the inner
disk profile is difficult to constrain due to its small physical scale
($\sim$0.1 au). Previous studies on dust traps created by changes in
the viscosity \citep[e.g.,][]{Kretke07,Kretke09,Kretke10} are based on a prescribed $\alpha$ profile without
considering the detailed physics of the MRI. 

IOPF Paper V solved for the 2D structure of the inner disk
assuming a steady-state, viscously heated $\alpha$-disk, and with
$\alpha$ determined self-consistently (via vertical averaging) from
considerations of the MRI and non-ideal MHD effects. \citet{Jankovic2021} and \citet{Jankovic2022} presented more detailed models of DZIBs, including effects of dust on the ionization fraction. \citet{cecil2024} have presented numerical simulations of DZIBs, including how they vary in response to accretion disk instabilities.

Our goal in this study is to examine how disk properties and their evolution, i.e., changes in disk structure caused by declining accretion rates, affect the trapping of pebbles at the DZIB, which is needed for the onset of IOPF.
In \S\ref{S:physics} we introduce the physical model of our evolving protoplanetary disk and discuss the basic physics of pebble radial drift and trapping. We then examine the particular conditions for trapping at the DZIB, especially constraints on the size distribution of pebbles, and discuss the mass fraction of pebbles that is expected to be trapped at this location. In \S\ref{S:conclusions} we summarize our conclusions and discuss the implications of these results for the onset of planet formation in the IOPF scenario, as well as the emergence of the transition disk phase of protoplanetary disks.

\section{Physical Model}\label{S:physics}

\subsection{Disk Structure}

In Paper IV we followed the evolution of pebbles from an outer scale of 30~au
to the inner region at 0.1~au, but used a disk model that assumed a constant value of the Shakura-Sunyaev $\alpha$ viscosity parameter. Now we include a change in viscosity
associated with the DZIB in the inner regions. Specifically, we utilize the
$\alpha$ profile around the DZIB derived in Paper V, here implemented
via polynomial fits.

\begin{figure*}
\centering
\includegraphics[width=1.0\textwidth]{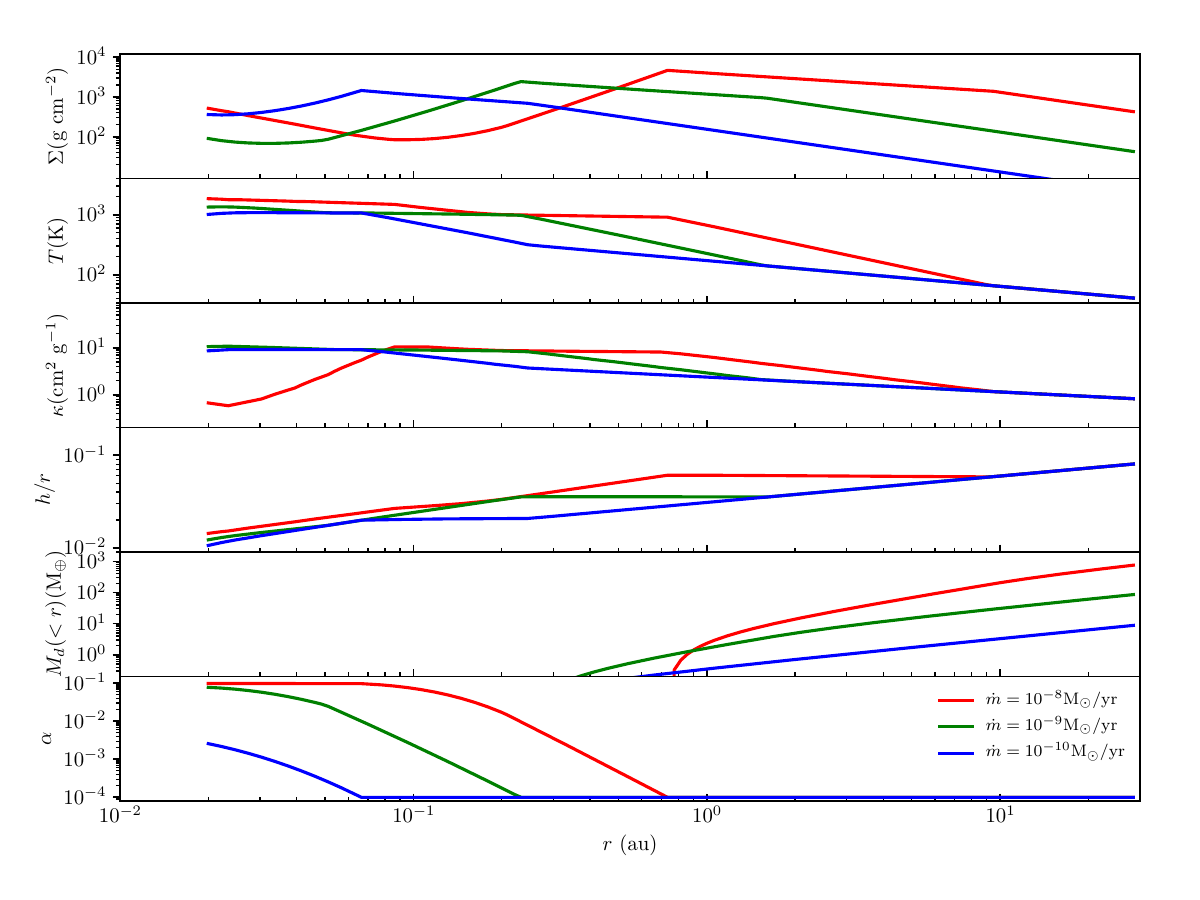}
\caption{
Structural profiles of Inside-Out Planet Formation (IOPF) disk models used in this paper, assuming
dead zone $\alpha=10^{-4}$ for accretion rates of $10^{-8}$, $10^{-9}$ and $10^{-10}\:M_\odot\:{\rm yr}^{-1}$. All models
are for a one solar mass central star. From top to bottom, the rows
show: gas mass surface density ($\Sigma_g$); midplane temperature
($T$); midplane opacity ($\kappa$) (assumed to be vertically
constant); disk aspect ratio ($h/r$); enclosed mass in solids
($M_d(<r)$), i.e., initially assumed to be dust, summed in the disk
from $r_{\rm 1200K}$ out to radius $r$ for a solid to gas mass ratio
of $f_s=0.01$; disk viscous $\alpha$ profile. Note the difference in
location and width of the $\alpha$ transition zones.}
\label{fig:disk}
\centering
\end{figure*}

The basic idea is to find out how evolving disk properties affect the ability of pebbles to be trapped at the DZIB leading to pebble
ring formation. As the disk evolves, i.e., as the accretion rate
declines, the location of the pressure maximum caused by the $\alpha$
transition associated with the DZIB moves inward. From Paper V, we have
three snapshots of the radial profiles of disk midplane $\alpha$ viscosity for the following accretion rates: 
$\dot{m}=10^{-8}\:M_\odot\:{\rm yr}^{-1}$, 
$\dot{m}=10^{-9}\:M_\odot\:{\rm
  yr}^{-1}$ and $\dot{m}=10^{-10}\:M_\odot\:{\rm
  yr}^{-1}$. These radial profiles are shown in the bottom panel of Figure~\ref{fig:disk}.
  
We next calculate other physical disk properties from 0.02 au to 30 au at 140
logarithmically spaced positions, following the same method as in Paper IV. The physical structure is solved first in the inner region as an active disk, i.e., dominated by viscous heating. Then we calculate a transition radius
where stellar irradiation heating dominates accretion heating. The
midplane temperatures of the active and passive disk models
\citep[based on][]{CG97} are compared, and the hotter one is chosen as
the actual value. The radial profiles of several physical quantities of the disk models, i.e., gas mass surface density ($\Sigma$), midplane temperature ($T$), midplane opacity ($\kappa$), disk aspect ratio ($h/r$), and enclosed mass in solids ($M_d(<r)$), are shown in the other panels of Figure~\ref{fig:disk}.

\begin{deluxetable}{ccccc}{H}
\label{tab:dzib}
\tablecaption{Model properties at and near the DZIB for disks with different accretion rates.}
\tablehead{$\dot{m}$ & $r_{\rm DZIB}$ & $\delta r_{\rm DZIB}$ & $v_{\rm r,g,DZIB}$ & $\frac{d \ln{P}^a}{d \ln{r}}$\\
($M_\odot\:{\rm yr}^{-1}$) & (au) & (au) & ($\rm cm\:s^{-1}$) & }
\startdata
$10^{-8}$ & 0.73 & 0.65 au & 2.35 & 1.14\\
$10^{-9}$ & 0.23 & 0.2 au & 1.38 & 0.935\\
$10^{-10}$ & 0.066 & $\geq$ 0.05 & 0.892 & 0.748
\enddata
\footnote{This is the maximum positive pressure gradient in the vicinity of the DZIB, i.e., just interior to the pressure maximum.}
\label{tab:dzib}
\end{deluxetable} 

The model properties in the vicinity of the DZIB are summarized in
Table~\ref{tab:dzib}. The region of the disk that is of particular interest for the trapping of pebbles is that just interior of the pressure maximum of the DZIB. As a result of the $\alpha$ transition, the pressure gradient is positive just inside of DZIB, so the gas has a super-Keplerian motion that acts to drive pebbles outwards. As $\alpha$ reaches a plateau in the innermost region, the pressure gradient reverts to its typical negative value, and we define $\delta r_{\rm DZIB}$ as the radial extent of the region with a positive pressure gradient. On the other hand, gas radial velocity inwards always tries to drag pebbles towards the star. Thus, very small particles that are well coupled to the gas always move inwards regardless of the variation of the pressure gradient. For a given disk structure, there is a minimum pebble size that will be trapped at the DZIB.

Note that as the accretion rate decreases, both the gas radial velocity and pressure gradient at the DZIB decrease. A smaller pressure gradient means pebbles have a smaller drift velocity with respect to the gas disk. The gas radial velocity also drops. This trend of reduced inward gas velocity makes it more difficult for gas drag to move pebbles inwards.
A detailed analysis of the trapping size and accretion rate will be presented in the following section. 


\subsection{Pebble Drift}

The radial drift of larger grains in protoplanetary disks can be
divided into two parts: the drift relative to gas and the imposed
inward drift due to viscous gas drag, i.e., dragged inwards by radial
gas flow. To describe these effects, we utilize the equation from
\citet{TL2002}:
\begin{eqnarray}
v_{r,p} &=&  v_{\rm drift}+v_{\rm drag} \nonumber \\
&\simeq&  -\frac{k_P(c_s/v_K)^2 }{\tau_{\rm fric}+\tau_{\rm fric}^{-1}} v_K+\frac{v_{r,g}}{1+\tau_{\rm fric}^2},
\label{eq:vrp}
\end{eqnarray}
where $c_s$ is the sound speed, $v_K$ is the local Keplerian orbital
velocity, $\tau_{\rm fric}\equiv \Omega_K t_{\rm fric}$ is the
dimensionless friction timescale and $\Omega_K$ is the orbital angular
frequency. For pebbles of a certain size, the other two parameters,
$k_P$ and $v_{r,g}$, constrain how the pebbles concentrate at a
pressure maximum. Here, $k_P$ is the power-law index of the pressure
radial profile, i.e., when expressed as $P=P_0(r/r_0)^{-k_P}$.
At the local pressure maximum of the DZIB, $k_P=0$, i.e., there is no
radial gradient of midplane gas pressure. Interior to the pressure maximum $k_P$ takes negative values. Finally, $v_{r,g}$ is the
radial velocity of gas due to viscous evolution.

In the inner part of the protoplanetary disk, the midplane temperature
is dominated by viscous heating powered by accretion, so the gas
radial velocity is (Paper I, eq.~12, note the differences in normalization of $\alpha$ and $r_{\rm au}$):
\begin{eqnarray}
v_{r,g} & = & - 3 \nu / (2 f_r r)\nonumber\\
|v_{r,g}| & = & \frac{3^{6/5}}{2^{12/5}\pi^{2/5}} \left(\frac{\mu}{\gamma k_B}\right)^{-4/5}
\left(\frac{\kappa}{\sigma_{\rm SB}}\right)^{1/5}\\ \nonumber
& \times & \alpha^{4/5}(G m_*)^{-1/5}f_r^{-3/5}\dot{m}^{2/5} r^{-2/5}\\
&\rightarrow & 1.93 \gamma_{1.4}^{4/5}\kappa_{10}^{1/5}\alpha_{-4}^{4/5}m_{*,1}^{-1/5} f_r^{-3/5} \dot{m}_{-9}^{2/5}r_{\rm 0.1 au}^{-2/5}\:{\rm cm\: s^{-1}},\nonumber
\end{eqnarray}
where $\gamma\equiv1.4\gamma_{1.4}$ with fiducial normalization to a
value of 1.4 for $\rm H_2$ with rotational modes excited, $\mu=2.33m_{\rm H}$ is the mean molecular mass, $\kappa
\equiv \kappa_{10}10\ {\rm cm^2\:g^{-1}}$ is the disk mean opacity
\citep[with fiducial normalization here appropriate for inner disk
conditions; but, note we will more generally use tabulated opacities
from][]{zhu2009opac},
$\sigma_{\rm SB}$ is the Stefan-Boltzmann constant, $m_*\equiv
m_{*,1}M_\odot$ is the stellar mass with fiducial normalization of
$1\:M_\odot$, $f_r \equiv1-\sqrt{r_*/r}$, $r_*$ is the stellar radius,
and $r_{\rm 0.1au}\equiv r/(0.1\:{\rm au})$. The gas radial velocity increases with the accretion rate $\dot{m}$ and decreases with radius $r$, with both scalings following a $2/5$ power-law dependence.

The disk aspect ratio, $h/r=c_s/v_K$, is an important metric of disk structure that can be related to pebble radial drift.
It can be written in the form (Paper III, eq.~30):
\begin{eqnarray}
\frac{h}{r} & = & \frac{c_s}{v_K}=\frac{3^{1/10}}{2^{7/10}\pi^{{1/5}}}\left(\frac{\mu}{\gamma k_B}\right)^{-2/5}\left(\frac{\kappa}{\sigma_{\rm SB}}\right)^{{1}/{10}}\\ \nonumber
&\times& \alpha^{-{1}/{10}}\left(Gm_*\right)^{-{7}/{20}}\left(f_r \dot{m}\right)^{1/5}r^{{1}/{20}}\\
&\rightarrow&0.037\gamma_{1.4}^{2/5}\kappa_{10}^{1/10}\alpha_{-4}^{-1/10}m_{*,1}^{-7/20}(f_r\dot{m}_{-9})^{1/5}{r_{\rm 0.1 au}^{1/20}}.\nonumber
\\\nonumber
\end{eqnarray}
The inner disk, where viscous heating sets the temperature structure, is quite thin, exhibiting a small flaring index of $0.05$.

\begin{figure*}[t]
\centering
\includegraphics[width=1.0\textwidth]{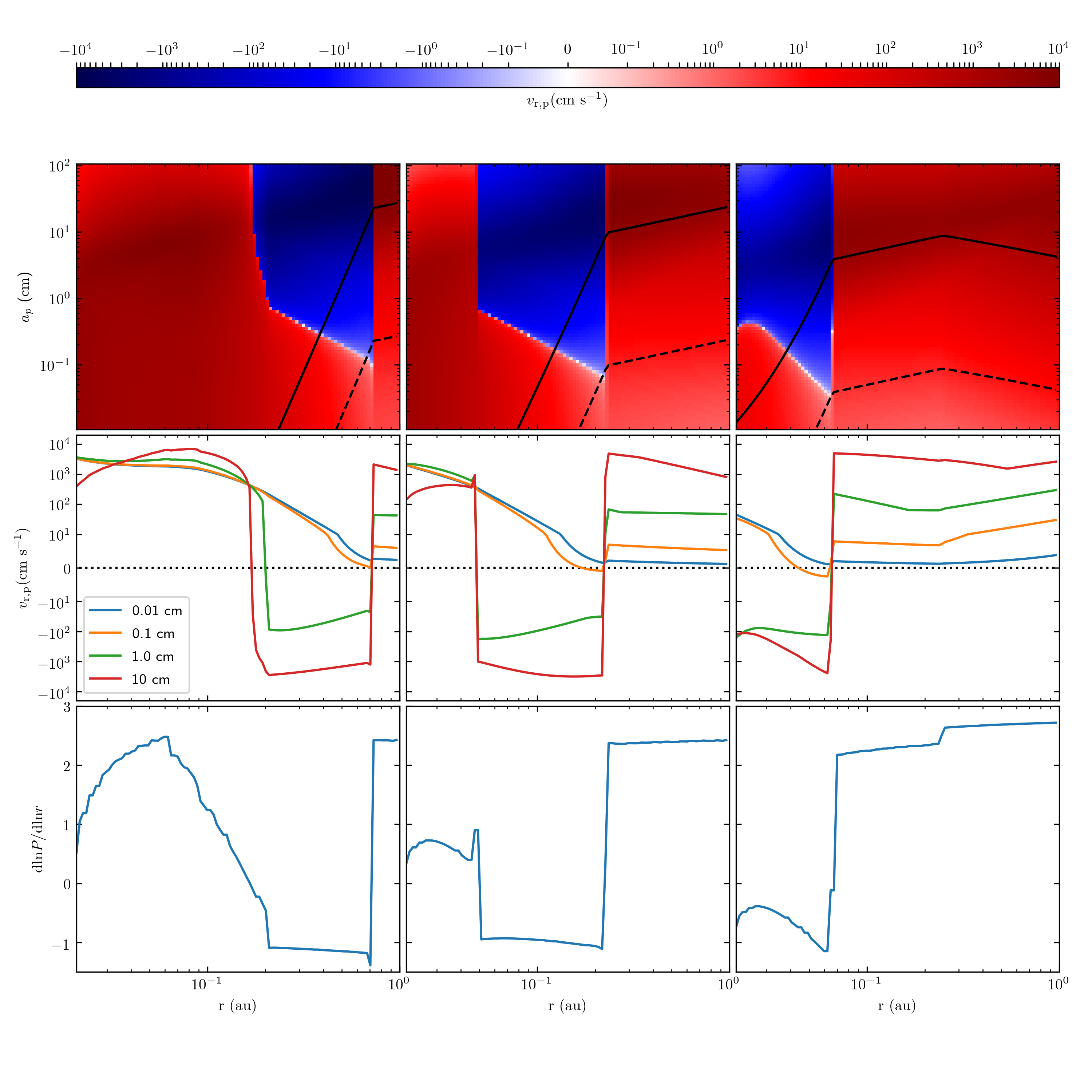}
\vspace{-0.5in}
\caption{
{\it (a) Top row:} Map of pebble radial drift velocity plotted in parameter space of disk radius, $r$, and 
pebble radial size, $a_p$. Red color is used for inward drift velocities; blue for outward velocities. From left to right, the panels show results for disks with 
$\dot{m}=10^{-8}\:M_\odot\:{\rm yr}^{-1}$, $\dot{m}=10^{-9}\:M_\odot\:{\rm yr}^{-1}$ and
$\dot{m}=10^{-10}\:M_\odot\:{\rm yr}^{-1}$. All disks have the same dead zone $\alpha=10^{-4}$. The minimum size of pebbles that experience a zone of outward migration (blue) and thus can be trapped near the DZIB becomes smaller as $\dot{m}$ decreases, i.e., 0.10 cm at $\dot{m}=10^{-8}\:M_\odot\:{\rm yr}^{-1}$, 0.063 cm at $\dot{m}=10^{-9}\:M_\odot\:{\rm yr}^{-1}$ and 0.035 cm at $\dot{m}=10^{-10}\:M_\odot\:{\rm yr}^{-1}$.
{\it (b) Middle row:} As (a), but now showing radial velocity of fixed-size pebbles as a function of radial location. 
{\it (c) Bottom row:} As (a), but now showing pressure gradient ($d \ln{P} / d \ln{r}  \equiv - k_P$) profile near DZIB. All models have similar values $k_P \simeq -1$ in the region just interior to the pressure maximum.}
\label{fig:vr_map1e-4}
\centering
\end{figure*}

\subsection{Pebble Trapping at the DZIB}

For the fastest-drifting pebbles ($\tau_{\rm fric}=1$) at 0.1~au, adopting the fiducial parameters such that $k_P = 51/20 = 2.55$ in a disk with constant $\alpha$ and $\kappa$, we obtain
$v_{\rm drift} \simeq 2.04\times10^4k_P\alpha_{-4}^{-1/5}{\rm cm\,s^{-1}}$ and
$v_{\rm drag} \simeq 0.97\alpha_{-4}^{4/5}{\rm cm\,s^{-1}}$.
Although these two characteristic velocity scales differ by several orders of magnitude, the drift velocity becomes negligible near pressure maxima where $k_P\simeq 0$.
As a consequence of the inward motion of the gas, the radial drift of pebbles must be directed outward relative to the gas flow, occurring in regions where $k_P$ is slightly negative.
Furthermore, the value of
$\alpha$ in this region is expected to be changing quite rapidly as
one leaves the dead zone. The opacities may also be varying, e.g., due
to the beginnings of dust destruction.
The assumption of $\tau_{\rm fric}=1$ will also be invalid for the
majority of pebbles, as gas densities are high and pebbles are more
coupled with gas flow. With the disk profiles and drift equations, we can numerically calculate the minimum pebble size to be trapped at DZIB.

In Figure~\ref{fig:vr_map1e-4} we show a ``pebble radial drift map'' near the DZIB, i.e., illustrating how pebble radial drift velocity varies with pebble size and radial location, for three different values of disk accretion rate, $\dot{m}$. In the bottom panels, the width of {\rm DZIB} region, i.e., from the pressure maximum to the location of maximum positive gradient, becomes much narrower in the low $\dot{m}$ cases. However, the pressure gradient remains at a similar level, i.e., $d \ln{P} / d \ln{r} \sim 1$. There are two factors at play: first, the pressure gradient is evaluated in logarithmic space of $P$ and $r$; second, the disk temperature always increases when $r$ gets smaller, neutralizing some of the effect of the jump in the magnitude of $\alpha$. In the following analysis, we will assume the value of $k_P$ stays constant with different $\dot{m}$ and values of dead zone $\alpha$. 
Figure~\ref{fig:vr_map1e-4} also shows radial profiles of $v_{r,p}$ for pebbles of fixed size across the DZIB regions. We see that the minimum size of pebbles to be trapped at the DZIB becomes
smaller with lower $\dot{m}$. However, the magnitude of the change is modest.
For the highest accretion rate, $\dot{m}=10^{-8}\:M_\odot\:{\rm yr}^{-1}$,
the minimum size is 0.10~cm. This size is 0.063~cm for 
$\dot{m}=10^{-9}\:M_\odot\:{\rm yr}^{-1}$ and 0.035~cm for
$\dot{m}=10^{-10}\:M_\odot\:{\rm yr}^{-1}$.

\begin{figure*}[t]
    \centering
    \includegraphics[width=0.45\textwidth]{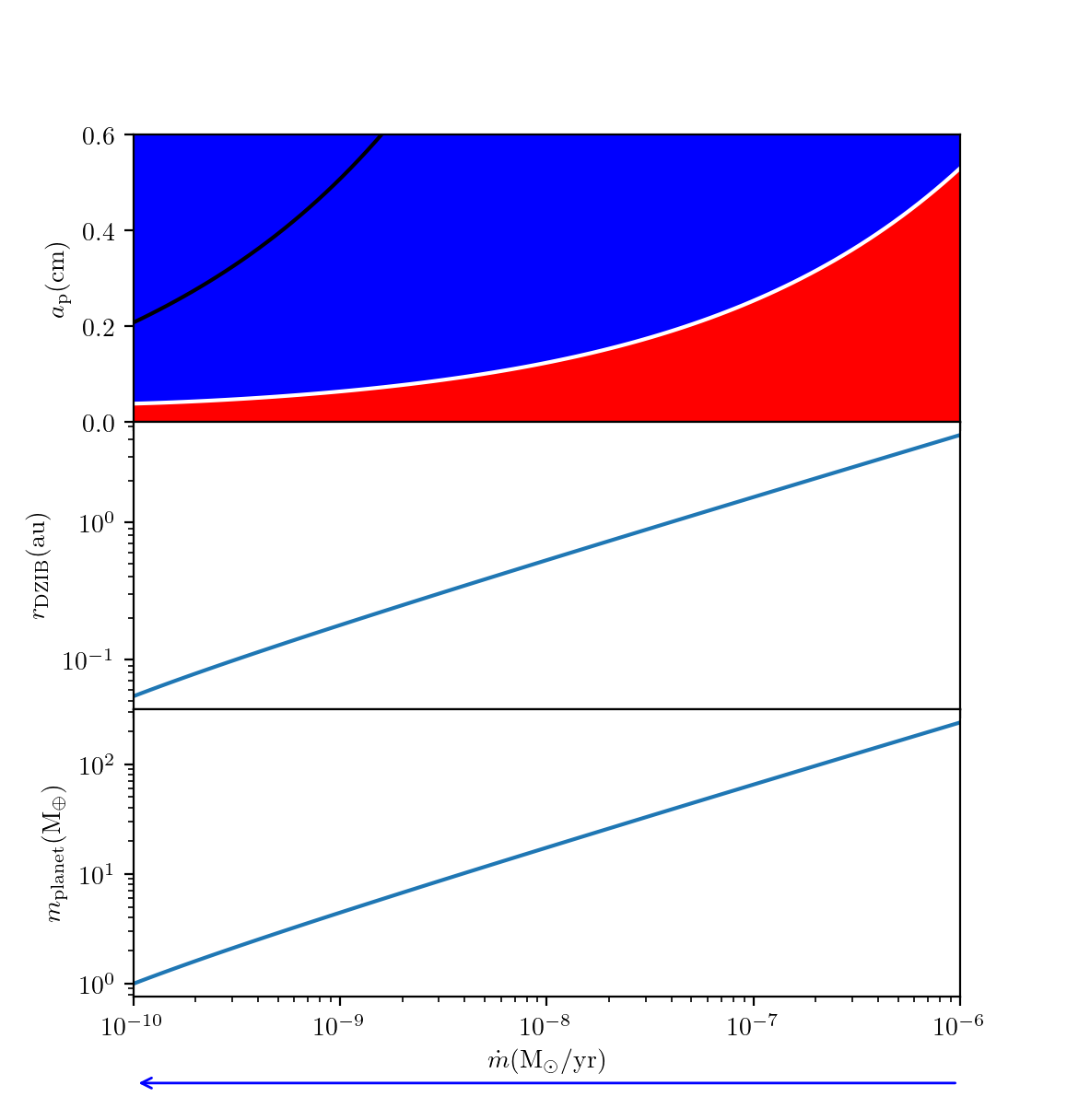}
    \includegraphics[width=0.45\textwidth]{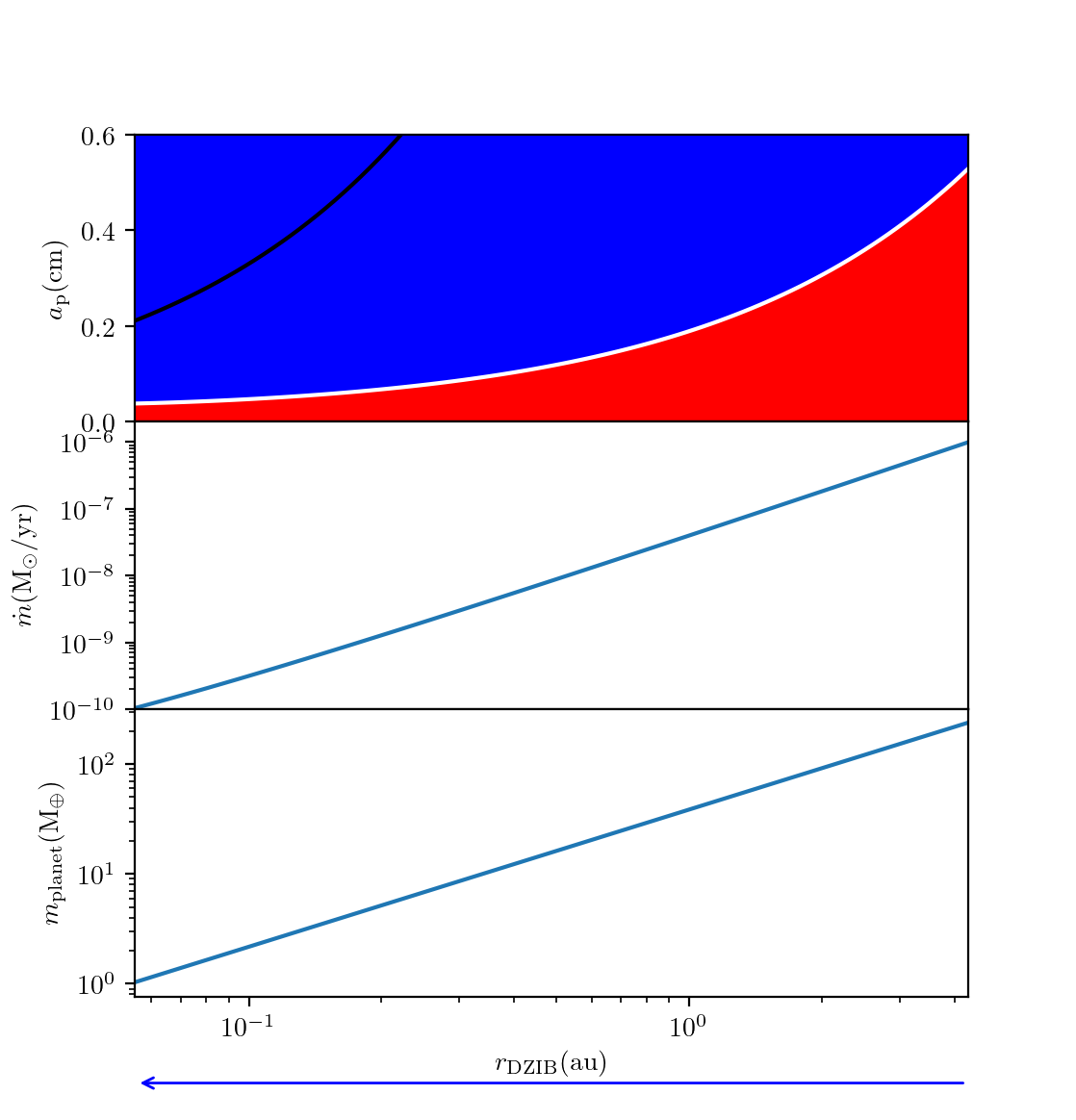}
    \caption{{\it (a) Left:} Evolution of disk and planet properties as a function of accretion rate. Note, the evolution during planet formation is expected to be from a high accretion rate state (on the right) to a low accretion rate state (on the left), as indicated by the arrow at the bottom of the figure. These results assume constant
    opacity $\kappa=10\:{\rm cm^2\: g}^{-1}$ and $\alpha=10^{-4}$.
    {\it Upper panel}: Minimum size for a pebble to be trapped at the
    DZIB versus accretion rate (white solid line). Pebbles in the blue region are trapped, while those in the red region are dragged inward through the DZIB along with gas accretion. The black solid line is the critical size to enter the Stokes drag regime (above the line) from the Epstein drag regime (below the line). 
    {\it Middle panel}: Location of pressure maximum predicted by Eq.~\ref{eq:r_dzib}. {\it Lower panel}: Mass of the first planet to form in the IOPF model due to shallow gap opening leading to pebble isolation (Eq.~\ref{eq:mg1}). 
    {\it (b) Right:} As (a), but now shown as a function of radial location of the DZIB, which evolves from large to small distances, and with the middle panel showing accretion rate versus $r_{\rm DZIB}$.
    } 
    \label{fig:trap_size}
\end{figure*}

From Equation~\ref{eq:vrp}, inside the pressure maximum of the DZIB, where gas orbits at super-Keplerian speeds, there is a competition between the tail wind induced outward radial drift and inward radial gas drag. For pebbles to be trapped, they need to be larger than a certain size to become decoupled from the inward radial gas drag and start to experience significant outward motion due to the super-Keplerian gas tail wind.
The location of the DZIB, i.e., where the disk reaches a temperature of about 1,200~K, is estimated as (Paper I, eq.~11 \& Paper IV, eq.~13):
\begin{equation}
    r_{\rm DZIB}=0.218\phi_{\rm DZIB}\gamma_{1.4}^{-2/9}\kappa_{10}^{2/9}\alpha_{-4}^{-2/9}m_{*,1}^{1/3}\left(f_r\dot{m}_{-9}\right)^{4/9}\ {\rm au},
    \label{eq:r_dzib}
\end{equation}
where $\phi_{\rm DZIB}$ is a dimensionless parameter of order unity accounting for potential differences from a pure viscous disk model. 

Thus, evaluating the condition for pebble trapping, i.e., $v_{r,p}>0$ at $r=r_{\rm DZIB}$, i.e., $|v_{\rm drift}|>|v_{\rm drag}|$, we find requires the following condition on pebble radius:
\begin{equation}
    a_{p,{\rm trap}} >0.0586\rho_{p,3}^{-1}k_P^{-1}\gamma_{1.4}^{-2/3}\kappa_{10}^{-1/3}\alpha_{-4}^{1/3}f_r^{-2/3}\dot{m}_{-9}^{1/3}\:{\rm cm},
    \label{eq:trap_size}
\end{equation}
where $\rho_{p,3}$ is the pebble density normalized to $3\:{\rm g\:cm}^{-3}$. Note, here we have adopted the Epstein drag regime. The midplane density at 0.1 au is (Paper I, eq.~2 and Paper III, eq.~31):
\begin{eqnarray}
\rho_g&=&2.52\times10^{-8}\gamma_{1.4}^{-6/5}\kappa_{10}^{-3/10}\alpha_{-4}^{-7/10}\nonumber\\
&\times&m_{*,1}^{11/20}(f_r\dot{m}_{-9})^{2/5}r_{\rm 0.1au}^{-33/20}\: \rm g\:cm^{-3}
\end{eqnarray}
The mean free path $\lambda =1/({n_{\rm H2}}{\sigma_{\rm H2}})$, with the number density $n_{\rm H2}=\rho_g/\mu$ and the cross section of the molecule ${\rm H_2}$ $\sigma_{\rm H2}\:=\:2\times10^{-15} \rm cm^2$, so we have the upper limit on the size of pebbles in the Epstein regime (Paper I, eq.~10):
\begin{eqnarray}
    \frac{9}{4}\lambda&=&0.173\gamma_{1.4}^{6/5}\kappa_{10}^{3/10}\alpha_{-4}^{7/10}\nonumber\\
    &\times&m_{*,1}^{-11/20}(f_r\dot{m}_{-9})^{-2/5}r_{\rm 0.1au}^{33/20}\: \rm cm
\end{eqnarray}
Note that this upper limit scales with $r^{33/20}$, so at $r_{\rm DZIB}$, it increases to $\sim0.6\:$cm, an order of magnitude greater than $a_{p,{\rm trap}}$. More complete comparisons between these two sizes are shown in the top panels of Figure \ref{fig:trap_size}.

Equation (\ref{eq:trap_size}) includes a weak dependence on opacity. However, this can be approximated as a constant, since it takes values $\kappa=7.93, 8.33, 9.11\:{\rm cm^2\:g^{-1}}$ when $\dot{m}=10^{-8}, 10^{-9}, 10^{-10}M_\odot\:{\rm yr}^{-1}$, respectively. The main result is that the minimum pebble size required for trapping at the DZIB shows a dependence on disk accretion rate as $a_{p,{\rm trap}}\propto \dot{m}^{1/3}$. Thus, as the disk evolves and the accretion rate declines, a larger and larger fraction of pebbles is expected to be trapped at the DZIB.

We note that the factor $f_r \equiv1-\sqrt{r_*/r}$ also varies with $\dot{m}$. However, for a typical value of $r_*\simeq 3\:R_\odot$, we find that $f_r = 0.86, 0.75, 0.54$ for the cases $\dot{m}=10^{-8}, 10^{-9}, 10^{-10}\:M_\odot\:{\rm yr}^{-1}$, so that $f_r^{-2/3} = 1.11, 1.21, 1.51$, so it has a weaker effect compared to that of the declining accretion rate. For simplicity, we adopt $f_r=1$ in the function for $r_{\rm DZIB}$, while $f_r \equiv1-\sqrt{r_*/r}$ for all other quantities at $r_{\rm DZIB}$.

The top panels of Figure~\ref{fig:trap_size} show the dependence of $a_{p,{\rm trap}}$ as a function of accretion rate and as a function of radius of the DZIB. Note, the accretion rate is expected to decline as the disk evolves and, associated with this, $r_{\rm DZIB}$ is also expected to decrease. These panels also show the size of pebbles that would be subject to the Stokes drag regime, thus validating the use of the Epstein regime in the derivation of Eq.~(\ref{eq:trap_size}). Other panels in Figure~\ref{fig:trap_size} show how accretion rate and $r_{\rm DZIB}$ vary with each other and how the expected planet mass formed at the DZIB in the IOPF model (set by shallow gap opening) varies with these quantities. We note that the precise definition of the planet mass is introduced later in Eq.~\ref{eq:mg1}.

\subsection{Pebble trapping fraction at the DZIB}

To evaluate the mass fraction of pebbles that are trapped at the DZIB, $f_{\rm trap}$, and the complementary leakage fraction, $f_{\rm leak}\equiv1-f_{\rm trap}$, we need to consider the size distribution of the pebbles. Grain and pebble growth processes, mainly occurring in the outer disk, are expected to lead to a general increase in pebble size as the disk evolves. For example, in IOPF Paper IV it was found that in the fiducial case of dust evolving via sweep-up growth and pebble-pebble collisions, the characteristic size of pebbles reaching the inner $\sim 1\:$au achieved (after $\sim 0.5\:$Myr of evolution) a fairly narrow distribution peaked near $\sim 1\:$cm in radius. This characteristic size declined to about 1~mm for locations in the disk around $\sim 5\:$au. These results were quite insensitive to the choice of $\alpha$, i.e., varied in the range from $10^{-4}$ to $10^{-3}$. 

Simulations with the DustPy code \citep{Stammler2022} of dust evolution within a fixed background gas model of the Paper IV disk, i.e., with $\dot{m}_{-9}=1$ and $\alpha=10^{-4}$, have been performed by \citet{Hjalt2024}, who also find that inside 1~au most of solids mass are dominated by the upper end of the pebble size, with a narrow distribution that is peaked at $\sim0.6\:$cm (for fragmentation velocity of 1~m/s). 

The above results suggest that as the disk evolves to lower accretion rates, leading to inward motion of the DZIB and reduction of the minimum size for pebble trapping that the trapping fraction could dramatically increase once this size coincides with that of the typical pebble size at the location of $r_{\rm DZIB}$.

\begin{figure*}[t]
    \centering
    \includegraphics[width=0.75\textwidth]{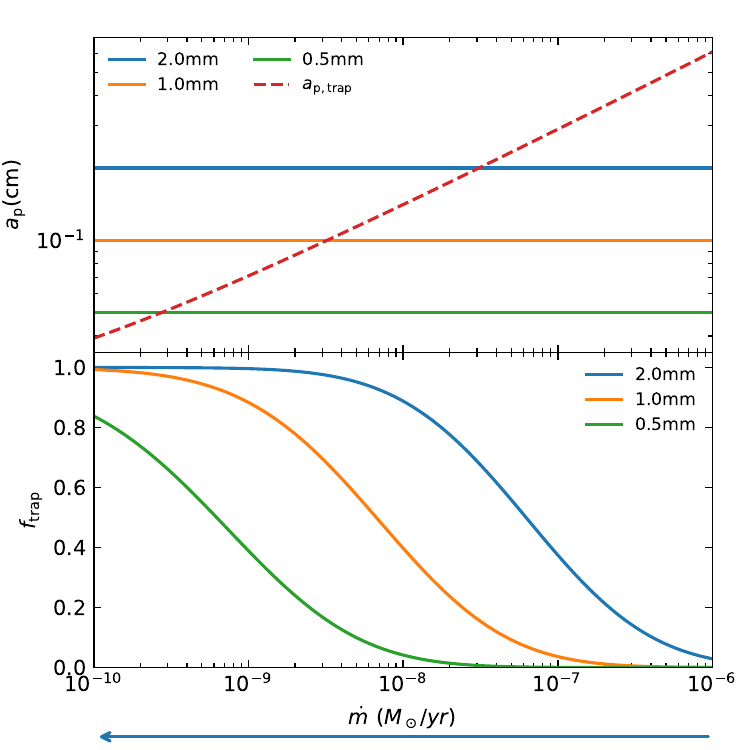}
    \caption{{\it (a) Top panel:} Minimum trapping size at the DZIB location (red dashed line), shown as a function of disk accretion rate, with evolution expected to occur from right to left. The peak pebble sizes of three assumed log-normal distributions (peaking at 0.5, 1, and 2 mm and with dispersion $\sigma_p=0.2$~dex) are indicated by the horizontal lines.
    {\it (b) Bottom panel:} Evolution of DZIB pebble trapping fraction, $f_{\rm trap}$, with disk accretion rate for the three pebble size distributions described in (a).  
    }
    \label{fig:leak_frag}
\end{figure*}

In Figure~\ref{fig:leak_frag}a we show the evolution of the minimum size of pebbles to be trapped at the DZIB, $a_{\rm trap}$, as a function of disk accretion rate. We also indicate the peak pebble sizes of three assumed log-normal distributions, peaking at 0.5, 1, and 2 mm and with dispersion $\sigma_p=0.2$~dex, which are guided by the results of IOPF Paper IV. Note that this log-normal distribution is intended to approximate the upper end of the
evolved pebble size distribution; the full width of the large-grain component that
dominates the solid mass is approximately 1~dex (corresponding to $5\sigma$). In Figure~\ref{fig:leak_frag}b, we then show the evolution of the resulting pebble trapping fraction at the DZIB, $f_{\rm trap}$. We see that this fraction rises from near zero during the early, high accretion rate stages, to approach unity by the final low accretion rate phase. The precise accretion rate at which this transition occurs depends on the assumed pebble size distribution. For pebble sizes of 0.5, 1, and 2~mm, the transition, defined by $f_{\rm trap}=0.5$, occurs at accretion rates of $\sim6\times 10^{-10}$, $6\times 10^{-9}$, and $6\times 10^{-8}\:M_\odot\:{\rm yr}^{-1}$, respectively. Keeping this log-normal size distribution, a smaller dispersion leads to a much sharper
transition in trapping efficiency over the course of disk evolution. For example, adopting
$\sigma = 0.1$~dex with a peak size of 1~mm yields a minimum size of $\sim 0.5$~mm, in which
case the trapping efficiency is nearly zero at $\dot{m}=10^{-8}\,M_\odot\,{\rm yr^{-1}}$ and
then rises rapidly to almost 100\% by $\dot{m}=10^{-9}\,M_\odot\,{\rm yr^{-1}}$. In contrast,
a broader distribution with $\sigma = 0.4$~dex corresponds to a ``full width'' spanning
approximately $0.05$--$20$~mm. This wider range produces a trapping efficiency of $\sim 15\%$
even at $\dot{m}=10^{-6}\,M_\odot\,{\rm yr^{-1}}$, followed by a much flatter increase that
does not approach unity until $\dot{m}\simeq 10^{-10}\,M_\odot\,{\rm yr^{-1}}$. We thus see that the onset of pebble trapping at the DZIB is quite sensitive to the pebble size distribution in this location, which thus motivates the need for improved studies of the processes of pebble growth and destruction in these environments. 

Here, we discuss some additional considerations related to the pebble size distribution in DZIB regions. One process that may set a maximum size of pebbles is collisional fragmentation. The turbulent fragmentation limit on pebble radius is \citep[][eq.8]{2012A&A...539A.148B}:
\begin{equation}
    a_{p,{\rm frag}}=f_{f}\frac{2}{3\pi}\frac{\Sigma}{\rho_p\alpha}\frac{u_f^2}{c_s^2},
    \label{eq:a_turb}
\end{equation}
where $u_f$ is the fragmentation velocity and $f_f$ is a dimensionless factor (with fiducial value 0.37) by which the representative maximum pebble size is smaller than the theoretical limit. Inserting conditions of the DZIB, we have the following dependence of the pebble fragmentation limit on accretion rate:
\begin{equation}
    a_{p,{\rm frag}}=12.9 \frac{f_{f, 0.37} u_{f,10}^2}{\rho_{p,3}\gamma_{1.4}^{5/3}}\kappa_{10}^{-8/15}\alpha_{-4}^{-5/3}(f_r\dot{m}_{-9})^{1/3}~{\rm cm}.
    \label{eq:frag_size}
\end{equation}
There are two parameters that the fragmentation limit is most sensitive to: fragmentation velocity, $u_{f}$, and turbulent viscosity parameter, $\alpha$. The former has a high uncertainty, ranging from $\sim1\:{\rm m~s}^{-1}$ to $\sim10\:{\rm m~s}^{-1}$, based on different pebble compositions and environments. The $\alpha$ parameter is usually set to be constant during disk evolution for simplicity and is also highly uncertain.
Overall, we see that the turbulent fragmentation size limit is generally much larger than the minimum trapping size. We note that, other factors being equal, it is expected to decrease as disk evolves as $\dot{m}^{1/3}$, which is the same scaling as $a_{p,{\rm trap}}$.

If turbulent-driven fragmentation is the main process setting the size limit of pebbles, then the pebble size distribution is expected to be reasonably well described by a power-law \citep[e.g.,][]{Birnstiel2015,Okuzumi2016}, i.e.,
\begin{equation}
    \d m(a_p)\propto a_p^{-5/8}\d a_p.
\end{equation}
Assuming this power law holds from a minimum grain size of $0.1~{\rm \mu m}$ up to the fragmentation limit, given the minimum trapping size, we can then evaluate the mass fraction of pebbles that is trapped at the DZIB. Under such assumptions we find that the trapping fraction does not evolve significantly with disk accretion rate, e.g., having values ranging from 0.17 to 0.3 for accretion rates between $\dot{m}=10^{-10}-10^{-6}\:M_\odot\:{\rm yr}^{-1}$ for the case with fragmentation velocity of 10~m/s. 
However, we note the sensitivity of this result to the assumed minimum grain size, as well as the overall assumption that pebbles have time to grow to become limited by the fragmentation barrier.

\subsection{Impact of planet formation time}

The planetary mass required for pebble isolation via shallow gap opening,
i.e., the mass scale of planets forming via IOPF, is (Paper IV, eq.~15):
\begin{eqnarray}
\label{eq:mg1}
M_{\rm G,D}&=&\phi_{\rm G,D}\sqrt{3\pi}m_*v_K^{-5/2}\alpha^{1/2}c_s^{5/2}\\
&=&\phi_{\rm G,D}\frac{3^{3/4}}{2^{7/4}}\left(\frac{\mu}{\gamma k_B}\right)^{-1}\left(\frac{\kappa}{\sigma_{\rm SB}}\right)^{{1}/{4}} \alpha^{1/4}\nonumber\\
&\times&G^{-7/8}m_*^{1/8}\left(f_r\dot{m}\right)^{1/2}r^{{1}/{8}},\nonumber\\
&\rightarrow& 3.86 \gamma_{1.4}\kappa_{10}^{1/4}\alpha_{-4}^{1/4}m_{*,1}^{1/8}(f_r\dot{m}_{-9})^{1/2}r_{\rm 0.1 AU}^{1/8}\: \rm M_\oplus. \nonumber
\end{eqnarray}
Then the mass of the first, i.e., ``Vulcan'', planet to form is evaluated by taking $r_{\rm DZIB}$ as
the planet's location. We obtain 
\begin{equation}
    M_{p,1}=4.25  \gamma_{1.4}^{35/36} \kappa_{10}^{5/18}\alpha_{-4}^{2/9}m_{*,1}^{1/6}(f_r\dot{m}_{-9})^{5/9}\: \rm M_\oplus.
\end{equation}
Note the differences from eq.~17 in Paper IV, where we replaced $r_{0.1,{\rm au}}$ with $r_{\rm DZIB}$. Because $M_{\rm G,D}$ is only weakly dependent on orbital radius, $M_{p,1}$ retains a similar scaling with $\dot{m}$. Thus, we see that the innermost planet's mass varies from
$M_{p,1} = 1.18 M_\oplus$ for
$\dot{m}=10^{-10}\:M_\odot\:{\rm yr}^{-1}$,
to $4.25 M_\oplus$ for
$\dot{m}=10^{-9}\:M_\odot\:{\rm yr}^{-1}$, to $15.3
M_\oplus$ for $\dot{m}=10^{-8}\:M_\odot\:{\rm yr}^{-1}$ and that the required gap opening mass is significantly larger in the earlier, higher accretion rate phases when the DZIB is further out.

In the models developed in Paper IV, the disk outer boundary is supplied with dust and pebbles given the steady state accretion rate and our adopted value of
$f_s=0.01$, and the fiducial pebble to dust mass ratio is also set to 0.01
for the injected solids. Due to sweep-up growth, it was found that most solids are incorporated into pebbles, i.e., $f_p = \dot{m}_p/\dot{m}_s\simeq
1$. Then the accretion limited pebble supply rate is:
\begin{eqnarray}
\dot{m}_p &=&10^{-11} f_p f_{s,0.01} \dot{m}_{-9}\: M_\odot\:{\rm yr}^{-1}\\
&=& 3.33\times 10^{-6}f_p f_{s,0.01} \dot{m}_{-9}\: M_{\oplus}\:{\rm yr}^{-1}.\nonumber 
\label{eq:stable_flux}
\end{eqnarray}
This implies the formation time of the first, Vulcan planet is
\begin{eqnarray}
    t_{\rm form,1} & = & \frac{M_{p,1}}{f_{\rm trap}\dot{m}_p}\nonumber \\ & =& 1.28\times10^6 \frac{\gamma_{1.4}^{35/36} \kappa_{10}^{5/18}}{f_{\rm trap}f_pf_{s,0.01}}\alpha_{-4}^{2/9}m_{*,1}^{1/6}\frac{f_r^{5/9}}{\dot{m}_{-9}^{4/9}}\:{\rm yr}\nonumber\\
    \label{eq:tform}
\end{eqnarray}
Thus, we see that even when the trapping fraction rises to unity the timescale to form the first planet can be $\sim 1\:$Myr, which is a characteristic timescale of disk accretion rate evolution. However, as discussed in Paper IV, there could be a phase of boosted pebble supply rate during early phases of disk evolution, triggered by the time when a significant population of macroscopic pebbles develops that is able to engage in efficient sweep-up growth of the small grain dust population. Typical boost factors in the pebble supply rate in the first few~$\times10^5\:$yr were found to be $f_{p,{\rm boost}}\sim10$. In the case of significant enhancements in pebble supply rates, the timescales implied by equation~(\ref{eq:tform}) should be regarded as upper limits.

\begin{figure*}[t]
    \centering
    \includegraphics[width=0.75\textwidth]{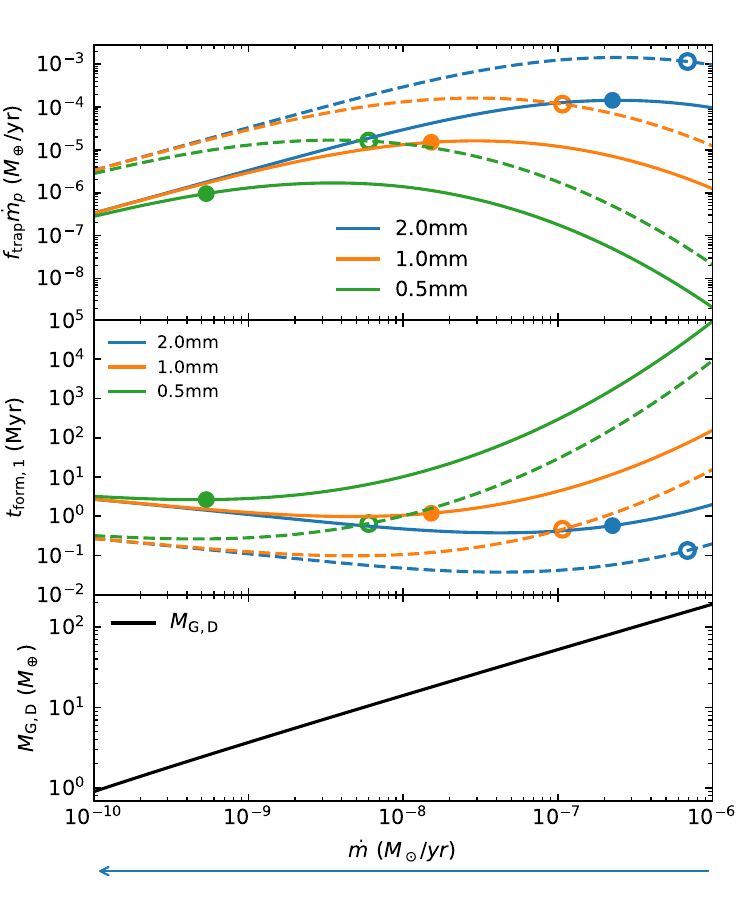}
    \caption{{\it (a) Upper panel:} Effective pebble flux that is trapped at the DZIB as a function of disk accretion rate for the models with $a_{p,{\rm peak}}=0.5, 1, 2\:$mm (green, orange, blue lines, respectively) and for $f_{p,{\rm boost}}=1$ and 10 (solid and dashed lines, respectively). For the fiducial evolving disk model in which accretion rate declines by a factor of 10 every 1~Myr, the point at which the time-integrated trapped pebble mass equals the gap opening mass is indicated by solid and open circles for cases with $f_{p,{\rm boost}}=1$ and 10, respectively.
    {\it (b) Middle panel:} Vulcan planet formation timescales, $t_{\rm form,1}\equiv M_{p,1}/(f_{\rm trap}\dot{m}_p)$, i.e., the time to accumulate enough material for a planet to have enough mass to open a shallow gap in the disk, i.e., with $M_{p,1}=M_{\rm G,D}$, for the same cases shown in (a). Note, these times are based on instantaneous rates. The solid and open circles mark the stages at which the time-integrated trapped pebble mass equals the gap opening mass, as described in (a).
    {\it (c) Bottom panel:} Evolution of gap opening mass, $M_{\rm G,D}$, at the DZIB, i.e., $M_{p,1}$, with disk accretion rate.
    }
    \label{fig:build_time}
\end{figure*}

The flux of trapped pebbles at the DZIB and local Vulcan planet formation time given this instantaneous pebble flux is shown in Figure~\ref{fig:build_time} for the three pebble size distributions ($a_{p,{\rm peak}}=0.5, 1, 2\:$mm), the fiducial accretion rate evolutionary model of IOPF Paper IV ($\dot{m}=\dot{m}_0e^{-t/t_0}$, with $t_0=4.34\times 10^5\:$yr (so that accretion rates decline by a factor of 10 every 1~Myr), and for cases with $f_{p,{\rm boost}}=1$ and 10.

As expected, we see that the flux of trapped pebbles is sensitive to the adopted pebble size distribution. For the case of $a_{p,{\rm peak}}=0.5$~mm it rises from very low values at early evolutionary phases, achieving a maximum value when $\dot{m}\lesssim10^{-8}\:M_\odot\:{\rm yr}^{-1}$. For the cases with larger pebble sizes, much higher fluxes of trapped pebbles are achieved at earlier stages.

Given the dependence of $M_{p,1}$ with accretion rate, the instantaneous formation times occupy a relatively narrow range, independent of the adopted pebble size distribution. For $f_{p,{\rm boost}}=1$, $t_{\rm form,1}\sim 1-3\:$Myr. For $f_{p,{\rm boost}}=10$, the formation times are reduced to $\sim0.1-1\:$Myr.

With an accretion rate that declines by an order of magnitude per Myr, both $r_{\rm DZIB}$ and the gap-opening mass $M_{p,1}$ evolve during the interval over which the planet accumulates most of its mass. Figure~\ref{fig:build_time} indicates (solid circles for $f_{p,{\rm boost}}=1$; open circles for $f_{p,{\rm boost}}=10$) the locations along the evolutionary tracks when the time integrated trapped pebble mass equals the required planetary gap opening mass at the current DZIB location. 

\begin{figure*}[t]
    \centering
    \includegraphics[width=1.0\textwidth]{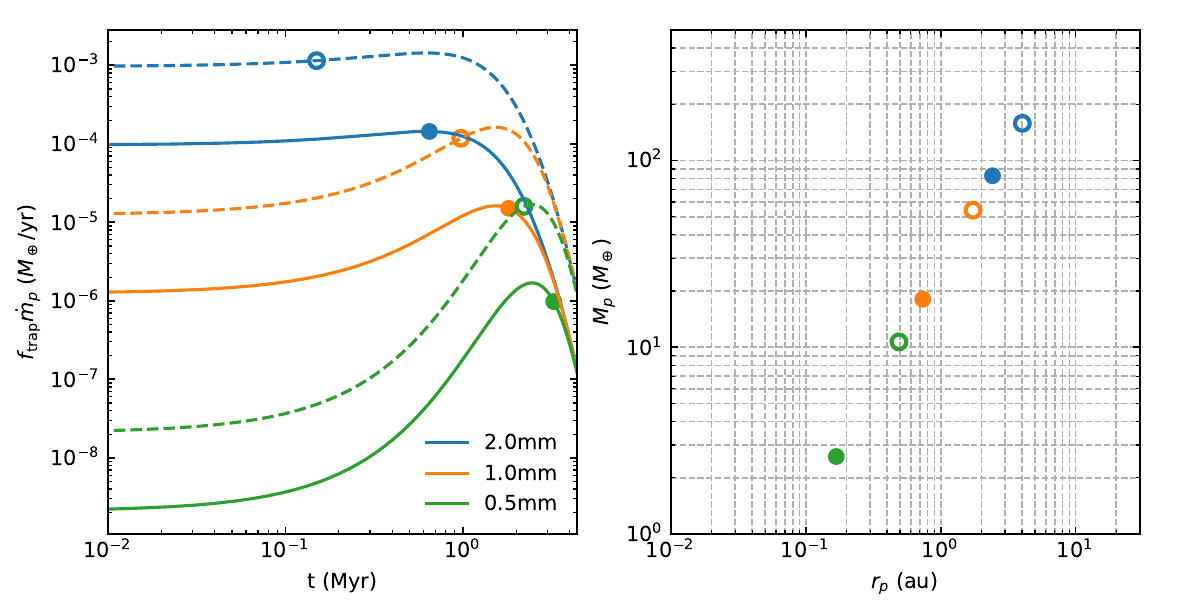}
    \caption{{\it (a) Left panel:} Time evolution of flux of trapped pebbles at the DZIB for the models with $a_{p,{\rm peak}}=0.5, 1, 2\:$mm (green, orange, blue lines, respectively) and for $f_{p,{\rm boost}}=1$ and 10 (solid and dashed lines, respectively).
    {\it (b) Right panel:} Mass versus semi-major axis for the Vulcan planets formed for the models shown in (a).
    }
    \label{fig:tform}
\end{figure*}

Figure~\ref{fig:tform}a shows the time evolution of flux of DZIB-trapped pebbles and times of formation of the Vulcan planets. Note, $t=0$ is defined here as the point in the disk evolution when the accretion rate is $10^{-6}\:M_\odot\:{\rm yr}^{-1}$. Figure~\ref{fig:tform}b shows the mass versus semi-major axis, i.e., DZIB location at time of gap opening, for the Vulcan planets. For example, for $a_{p,{\rm peak}}=0.5$~mm and $f_{p,{\rm boost}}=1$, the Vulcan planet forms with about $3M_\oplus$ at a location of about $0.2\:$au. Setting $f_{p,{\rm boost}}=10$ causes this planet to form with $\sim10M_\oplus$ at about 0.5~au. Larger values of $a_{p,{\rm peak}}$ lead to more massive Vulcan planets at semi-major axes $\gtrsim1\:$au. 

From these results, we conclude that for the model of Inside-Out Planet Formation presented here to reproduce observed properties of STIPs, i.e., the masses and semi-major axes of their Vulcan (innermost) planets, requires typical pebble sizes of $\sim0.5\:$mm at the location of the DZIB. We note that processes that we have not accounted for in our model, such as DZIB fluctuations induced on local scales, e.g., by turbulence, or global scales, e.g., by accretion rate fluctuations, including bursts, may lead to reduced trapping efficiency of pebbles at the DZIB. In this case, onset of planet formation at locations close to 0.1~au may require larger pebble sizes. 

The results discussed above may be influenced by the assumed evolution of the disk accretion rate, motivating a brief consideration of alternative $\dot{m}(t)$ histories. Our fiducial exponential decay model, in which $\dot{m}$ decreases by an order of magnitude per Myr, is motivated by Paper~IV and is broadly consistent with observationally inferred disk lifetimes of $\sim 1$--$10\ {\rm Myr}$. A slower decline in $\dot{m}$ would keep the DZIB at larger radii for longer and increase the associated gap-opening mass, thereby lengthening the formation time of the Vulcan planet and producing a more massive planetary system with generally wider orbital spacing. In more extreme cases, the system may remain in a high-accretion, low-trapping-efficiency state for an extended period, effectively stalling the onset of IOPF and yielding longer formation times and fewer planets. Conversely, a more rapid decay of $\dot{m}$ or early depletion of the inward pebble flux could inhibit the growth of the first planet before it reaches the shallow gap-opening mass to trigger DZIB retreat, thereby halting the subsequent IOPF process. However, it remains uncertain whether outer dust traps can form before the development of the DZIB. Rings and gaps are ubiquitous features in gaseous disks, yet their formation mechanisms—and consequently their characteristic timescales—are still not well constrained. Moreover, recent theoretical studies suggest that these substructures behave more like dust ``traffic jams'' rather than efficient dust traps \citep[e.g.,][]{hu2022}. Such rapid $\dot{m}$ declines are also in tension with viscous evolution models, as they imply unphysically fast depletion of the gas disk mass. While wind-driven accretion can remove gas more efficiently, it does not substantially alter the midplane dust reservoir, and a detailed examination of pebble delivery in wind-driven disks is beyond the scope of this work.

\section{Conclusions and Discussion}\label{S:conclusions}

When gas radial drag on pebbles is considered, the pressure bump in the inner region of a protoplanetary disk is not a pure particle trap anymore. We first examined disk properties at three accretion rates-$10^{-8}$, $10^{-9}$, and $10^{-10}~M_\odot~{\rm yr^{-1}}$-using realistic $\alpha$-viscosity profiles derived from models of the ionization fraction of the inner disk (IOPF Paper V). We found that the critical condition for pebble trapping, namely the pressure gradient at the DZIB that controls the outward drift velocity, remains approximately constant across these distinct accretion rates. Motivated by this result, we adopted analytic models that span a broader range of accretion rates and ultimately incorporate long-term disk evolution. The minimum trapping size, $a_{p,{\rm trap}}$ decreases modestly as the accretion rate declines. 

To better isolate the effects of pebble size and disk evolution, we approximated the pebble size distribution near its upper end using a narrow log-normal distribution (full width $\simeq1$~dex), which corresponds to the size range where most of the solid mass is concentrated. We find that the trapped pebble flux varies significantly with pebble size, and in order to form a $\sim3M_\oplus$ planet near 0.1 au, the majority of the pebble mass must be concentrated around 0.5~mm.

In addition to the pebble size at the DZIB, the primary uncertainty in our model lies in the disk conditions at this location. Specifically, dust dynamics just outside the active zone likely involve more than simple radial drift balanced by gas drag. Meridional flows are known to lift dust particles from the midplane into the disk surface \citep{hu2022}, where smaller grains may be entrained and ejected by disk winds. The accretion flow structure is also complex: simulations incorporating various ionization prescriptions have demonstrated that accretion can be dominated either by surface layers \citep{wbg2019,hu2021} or concentrated near the midplane \citep{hu2022}. An enhanced radial gas velocity, $v_{r,g}$, at the midplane significantly increases the critical trapping size, thereby favoring larger pebbles under the conditions relevant to the IOPF scenario. DZIB fluctuations due to local turbulence and/or accretion rate variations may also reduce the trapping efficiency, which would then require larger pebbles for the onset of first planet formation. Future work on realistic simulations of DZIB conditions coupled with dust and pebble evolution are needed to explore these possibilities.

Observationally, the onset of IOPF is expected to lead to a reduction in the amount of dust and pebbles in the very inner disk. Such systems are then expected to appear as ``transition disks'' in which the global spectral energy distribution lacks emission from dust at these locations \citep[e.g.,][]{manara2022,vandermarel2023}. IOPF and the model of pebble trapping we have presented here may provide an explanation for the emergence of the transition disk phase at accretion rates $\sim 10^{-9}\:M_\odot\:{\rm yr}^{-1}$ \citep[e.g.,][]{manara2014}.

\begin{acknowledgements}
We thank the anonymous referee for insightful and constructive comments, which significantly improved the quality of this manuscript. We also thank Til Birnstiel and Markus Hj\"alt for helpful discussions.
\end{acknowledgements}

\end{CJK*}

\begin{thebibliography}

\bibitem[Balbus \& Hawley(1991)]{balbus91}
Balbus, S. A. \& Hawley, J. F. 1991, \apj, 376, 214

\bibitem[Banzatti et al.(2022)]{Banzatti22}
Banzatti, A., Pontoppidan, K. M., Carr, J. S. et al. 2023, ApJL, 957, L22

\bibitem[Baruteau et al.(2014)]{2014prpl.conf..667B} Baruteau, C., Crida, A., Paardekooper, S.-J., et al.\ 2014, Protostars and Planets VI, 667. 

\bibitem[Baruteau \& Zhu(2016)]{BZ2016} 
Baruteau, C., \& Zhu, Z.\ 2016, \mnras, 458, 3927 

\bibitem[Birnstiel et al.(2012)]{2012A&A...539A.148B} Birnstiel, T., Klahr, H., \& Ercolano, B.\ 2012, \aap, 539, A148. 

\bibitem[Birnstiel et al.(2015)]{Birnstiel2015} 
Birnstiel, T., Andrews, S.~M., Pinilla, P., \& Kama, M.\ 2015, \apjl, 813, L14 

\bibitem[Bitsch et al.(2015)]{Bitsch2015}
Bitsch, B., Lambrechts, M. \& Johansen, A. 2015, A\&A, 582, 112

\bibitem[{Cecil} \& {Flock}(2024)]{cecil2024}
Cecil, M. \& Flock, M. 2024, A\&A, 692, 171

\bibitem[Cevallos Soto et al.(2022)]{CevallosSoto22}
Cevallos Soto, A., Tan, J. C., Hu, X., Hsu, C.-J., Walsh, C. 2022, \mnras, 517, 2285 (IOPF Paper VII)


\bibitem[Chiang \& Goldreich(1997)]{CG97}
 Chiang, E.~I., \& Goldreich, P.\ 1997, \apj, 490, 368 

\bibitem[{Chiang} \& {Laughlin}(2013)]{2013MNRAS.431.3444C}
{Chiang}, E., \& {Laughlin}, G. 2013, \textit{MNRAS}, 431, 3444

\bibitem[{Chatterjee} \& {Tan}(2014)]{CT14}
Chatterjee, S. \& Tan, J. C. 2014, ApJ, 780, 53

\bibitem[{Cossou} {et~al.}(2014)]{2014IAUS..299..360C}
{Cossou}, C., {Raymond}, S.~N. \& {Pierens}, A. 2014, \textit{IAUS}, 299, 360C

\bibitem[{Cossou} {et~al.}(2013)]{2013A&A...553L...2C}
{Cossou}, C., {Raymond}, S.~N. \& {Pierens}, A. 2013, \textit{A\&A}, 553L, 2C

\bibitem[Dong et al.(2017)]{Dong2017}
Dong, R., Li, S., Chiang, E., \& Li, H.\ 2017, \apj, 843, 127 

\bibitem[Emsenhuber et al.(2021)]{Emsenhuber2021}
Emsenhuber, A., Mordasini, C., Burn, R., Alibert, Y., Benz, W., Asphaug, E. 2021, A\&A, 656, 69

 
\bibitem[Fabrycky et al.(2014)]{2014ApJ...790..146F} Fabrycky, D.~C., Lissauer, J.~J., Ragozzine, D., et al.\ 2014, \apj, 790, 146. doi:10.1088/0004-637X/790/2/146

\bibitem[{Hansen} \& {Murray}(2013)]{2013ApJ.775.53}
{Hansen}, B. \& {Murray}, N. 2013, \textit{ApJ}, 775, 53

\bibitem[{Hansen} \& {Murray}(2012)]{2012ApJ.751.158}
{Hansen}, B. \& {Murray}, N. 2012, \textit{ApJ}, 751, 158

\bibitem[{Hj\"alt}(2024)]{Hjalt2024}
Hj\"alt, M. 2024, Master's thesis in Physics, Chalmers University of Technology (Examiner: Prof. J. C. Tan)

\bibitem[Hu et al.(2016)]{Hu2016}
Hu, X., Zhu, Z., Tan, J.~C. \& Chatterjee, S. 2016, \apj, 816, 19 (IOPF Paper III)

\bibitem[Hu et al.(2018)]{Hu2018} 
Hu, X., Tan, J.~C., Zhu, Z., et al.\ 2018, \apj, 857, 20 (IOPF Paper IV)

\bibitem[Hu et al.(2021)]{hu2021} Hu, X., Wang, L., Okuzumi, S., et al.\ 2021, \apj, 913, 2, 133 

\bibitem[Hu et al.(2022)]{hu2022} Hu, X., Li, Z.-Y., Zhu, Z., et al.\ 2022, \mnras, 516, 2, 2006

\bibitem[Jankovic et al.(2021)]{Jankovic2021}
Jankovic, M., Owen, J. E., Mohanty, S., Tan, J. C. 2021, MNRAS, 504, 280

\bibitem[Jankovic et al.(2022)]{Jankovic2022}
Jankovic, M., Owen, J. E., Mohanty, S., Tan, J. C. 2022, MNRAS, 509, 5974

\bibitem[{{Kley} \& {Nelson}(2012)}]{2012ARA&A..50..211K}
{Kley}, W. \& {Nelson}, R.~P. 2012, \textit{ARA\&A}, 50, 211

\bibitem[Kretke \& Lin(2007)]{Kretke07}
Kretke, K. \& Lin, D. N. C. 2007, ApJ, 664, L55

\bibitem[Kretke \& Lin(2010)]{Kretke10}
Kretke, K. \& Lin, D. N. C. 2010, ApJ, 721, 1585

\bibitem[Kretke et al.(2009)]{Kretke09}
Kretke, K. A., Lin, D. N. C., Garaud, P., Turner, N. J. 2009, ApJ, 690, 407

\bibitem[Manara et al.(2014)]{manara2014}
Manara, C. et al. 2014, A\&A, 568, A18


\bibitem[Manara et al.(2022)]{manara2022}
Manara, C. et al. 2022, Protostars and Planets VII, ASP Conf. Series, Vol. 534, Eds. S. Inutsuka, Y. Aikawa, T. Muto, K. Tomida, and M. Tamura. San Francisco: Astronomical Society of the Pacific, 2023, p.539

\bibitem[{{McNeil} \& {Nelson}(2010)}]{2010MNRAS.401.1691M}
{McNeil}, D.~S. \& {Nelson}, R.~P. 2010, \textit{MNRAS}, 401, 1691

\bibitem[Mohanty et al.(2018)]{Mohanty2018}
Mohanty, S., Jankovic, M.~R., Tan, J.~C., \& Owen, J.~E.\ 2018, ApJ, 861, 144 (IOPF Paper V)

\bibitem[{Ogihara} {et~al.}(2015)]{2015A&A...578A..36O}
 {Ogihara}, M., {Morbidelli}, A. \& {Guillot}, T. 2015, \textit{A\&A}, 578A, 36O

\bibitem[Okuzumi et al.(2016)]{Okuzumi2016} Okuzumi, S., Momose, M., Sirono, S., et al. 2016, \apj, 821, 2, 82. 

\bibitem[Pinilla et al.(2016)]{Pinilla2016}
 Pinilla, P., Klarmann, L., Birnstiel, T., et al.\ 2016, \aap, 585, A35 
 
\bibitem[Pinilla et al.(2016)]{Pinilla2016b}
 Pinilla, P., Flock, M., Ovelar, M.~d.~J., \& Birnstiel, T.\ 2016, \aap, 596, A81 

\bibitem[{Raymond} \& {Cossou}(2014)]{2014MNRAS.440L..11R}
{Raymond}, S.~N. \& {Cossou}, C. 2014, \textit{MNRAS}, 440L, 11R

\bibitem[{Rogers} \& {Owen}(2021)]{Rogers2021}
Rogers, J. G. \& Owen, J. E. 2021, MNRAS, 503, 1526

\bibitem[{Schlichting}(2014)]{2014ApJ...795L..15S}
{Schlichting}, H.~E. 2014, \textit{ApJ}, 795L, 15S

\bibitem[{Stammler} \& {Birnstiel}(2022)]{Stammler2022}
Stammler, S. \& Birnstiel, T. 2022, \apj, 935, 35

\bibitem[Takeuchi \& Lin(2002)] {TL2002}
Takeuchi, T., \& Lin, D.~N.~C.\ 2002, \apj, 581, 1344 

\bibitem[van der Marel(2023)]{vandermarel2023}
van der Marel, N. 2023, The European Physical Journal Plus, Volume 138, Issue 3, article id.225

\bibitem[Wang et al.(2019)]{wbg2019} Wang, L., Bai, X.-N., \& Goodman, J.\ 2019, \apj, 874, 1, 90

\bibitem[Zhu et al.(2009)]{zhu2009opac}
 Zhu, Z., Hartmann, L., \& Gammie, C.\ 2009, \apj, 694, 1045 

\bibitem[Zhu et al.(2012)]{zhu2012} 
Zhu, Z., Nelson, R.~P., Dong, R., Espaillat, C., \& Hartmann, L.\ 2012, \apj, 755, 6 


\end{thebibliography}
\end{document}